\documentclass[slac_one]{revtex4}
\usepackage{graphicx}
\usepackage{fancyhdr}
\usepackage{xspace}
\begin{document}

\title{Beyond the Standard Model Neutral Higgs Searches
at the Tevatron} 


%

\author{N. Krumnack on behalf of the CDF and D0 collaborations}
\affiliation{Baylor University, Waco, TX 76798, USA}

\begin{abstract}
We present an overview of the full range of Higgs searches in models beyond the Standard Model at the Tevatron.  This includes both searches for Fermiophobic Higgs and for SUSY Higgs at high $\tan\beta$.  No excess is seen in the data, so model dependent limits are set.
\end{abstract}

\maketitle

\thispagestyle{fancy}

\def\widtha{170mm}
\def\widthb{85mm}
\def\widthc{56.6667mm}
\def\widthcb{113.3333mm}
\def\widthd{42.5mm}
\def\widthdb{85mm}
\def\DO{D0\xspace}
\def\WW{$H\rightarrow WW$\xspace}
\def\WWW{$WH\rightarrow WWW$\xspace}
\def\dipho{$H\rightarrow \gamma\gamma$\xspace}
\def\tripho{$W^*\rightarrow h_fH\rightarrow h_fh_fW\rightarrow \gamma\gamma\gamma(\gamma)X$\xspace}
\def\bbb{$bH\rightarrow bbb$\xspace}
\def\ditau{$H\rightarrow\tau\tau$\xspace}
\def\bditau{$bH\rightarrow b\tau\tau$\xspace}

\section{Introduction}

Most of the Higgs efforts of the CDF and \DO collaborations are centered on finding a Standard Model Higgs boson.  However the Standard Model is incomplete and there are proposed extensions to the Standard Model, many of which feature one or more Higgs bosons as well.  These are often similar to the Standard Model Higgs, but differ in their couplings to other particles.  For the purposes of this presentation, I divide the Higgs searches into two categories, Fermiophobic Higgs searches and SUSY Higgs searches at high $\tan\beta$.

\section{Fermiophobic Higgs Searches}

In the Fermiophobic Higgs models, the Higgs is assumed not to couple to fermions.  At high masses it then decays to a W pair and at low masses to a photon pair.  The high mass searches are done in the channel \WWW, searching for a Higgs boson produced in association with a W boson.  This is done in order to avoid the higher backgrounds associated with the \WW channel.

\begin{figure*}[tbph]
\centering
\includegraphics[width=\widthd]{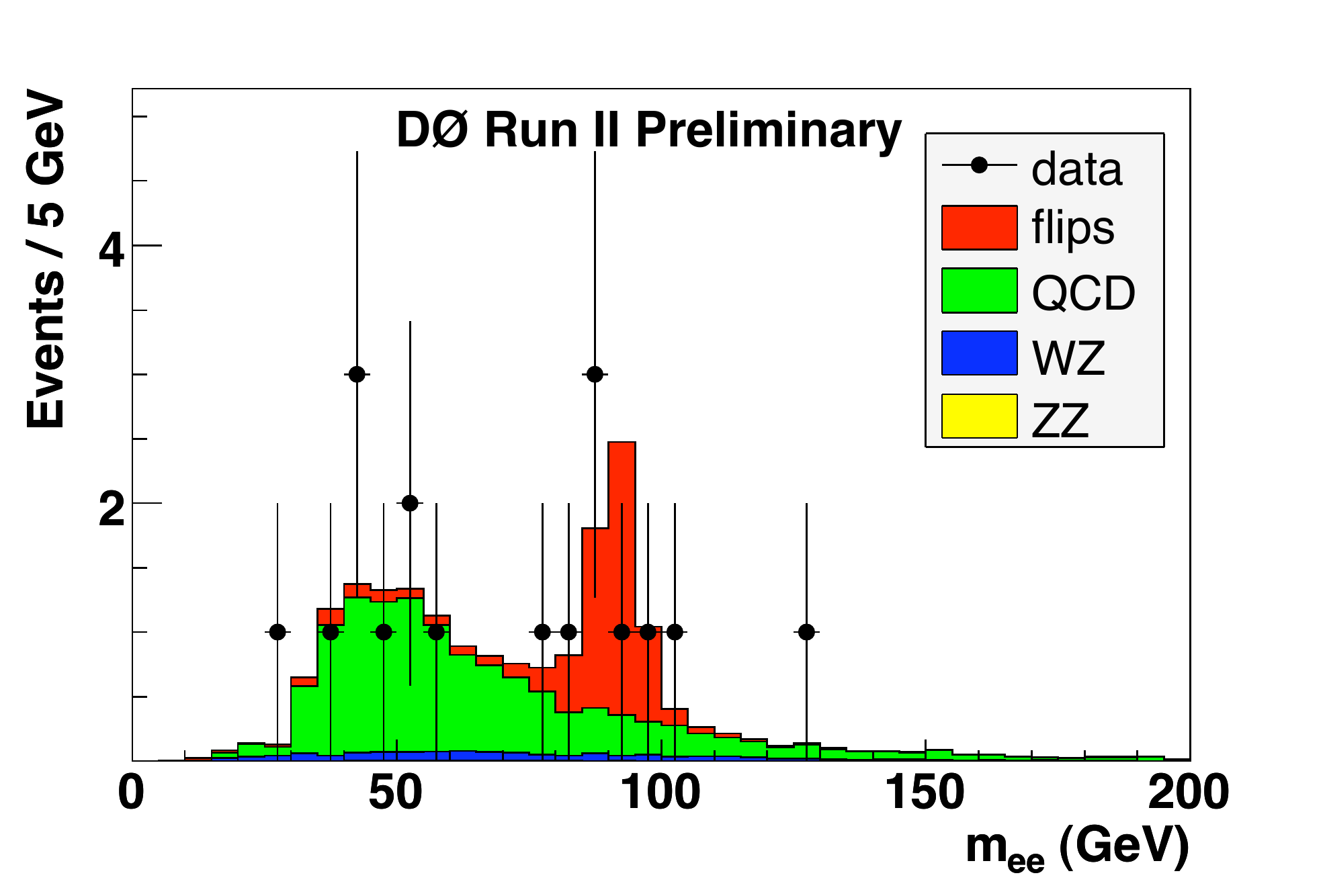}
\includegraphics[width=\widthd]{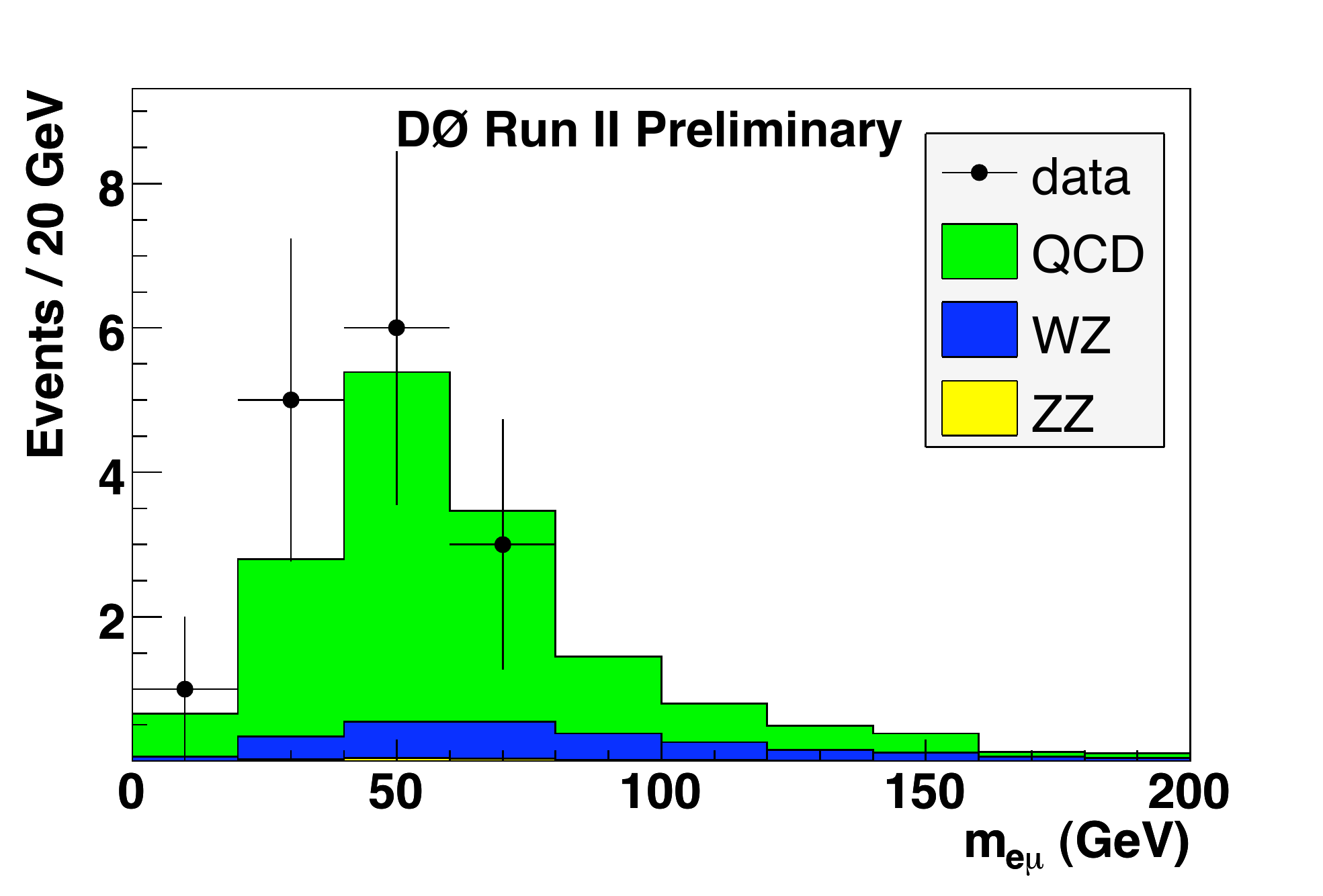}
\includegraphics[width=\widthd]{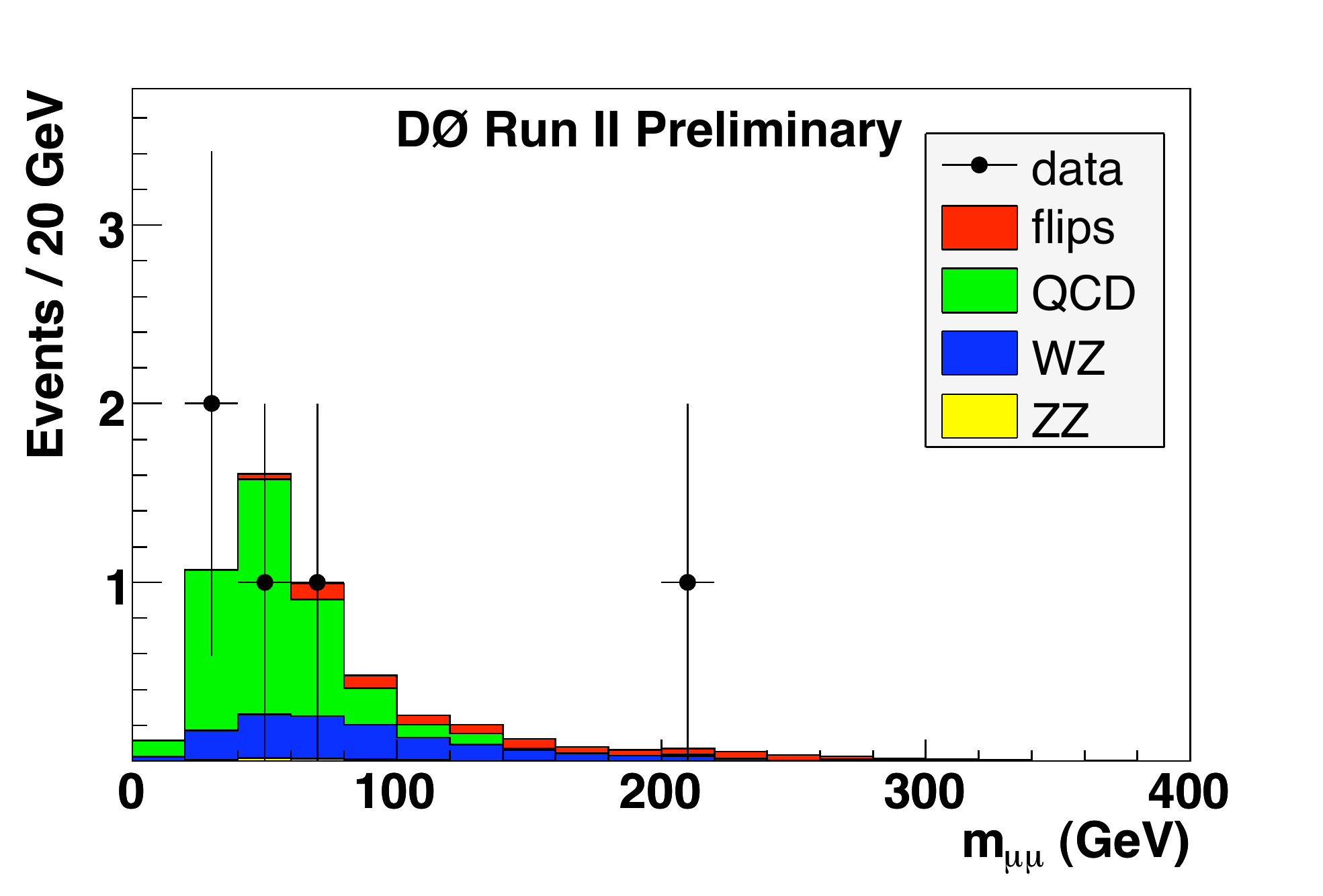}
\includegraphics[width=\widthd]{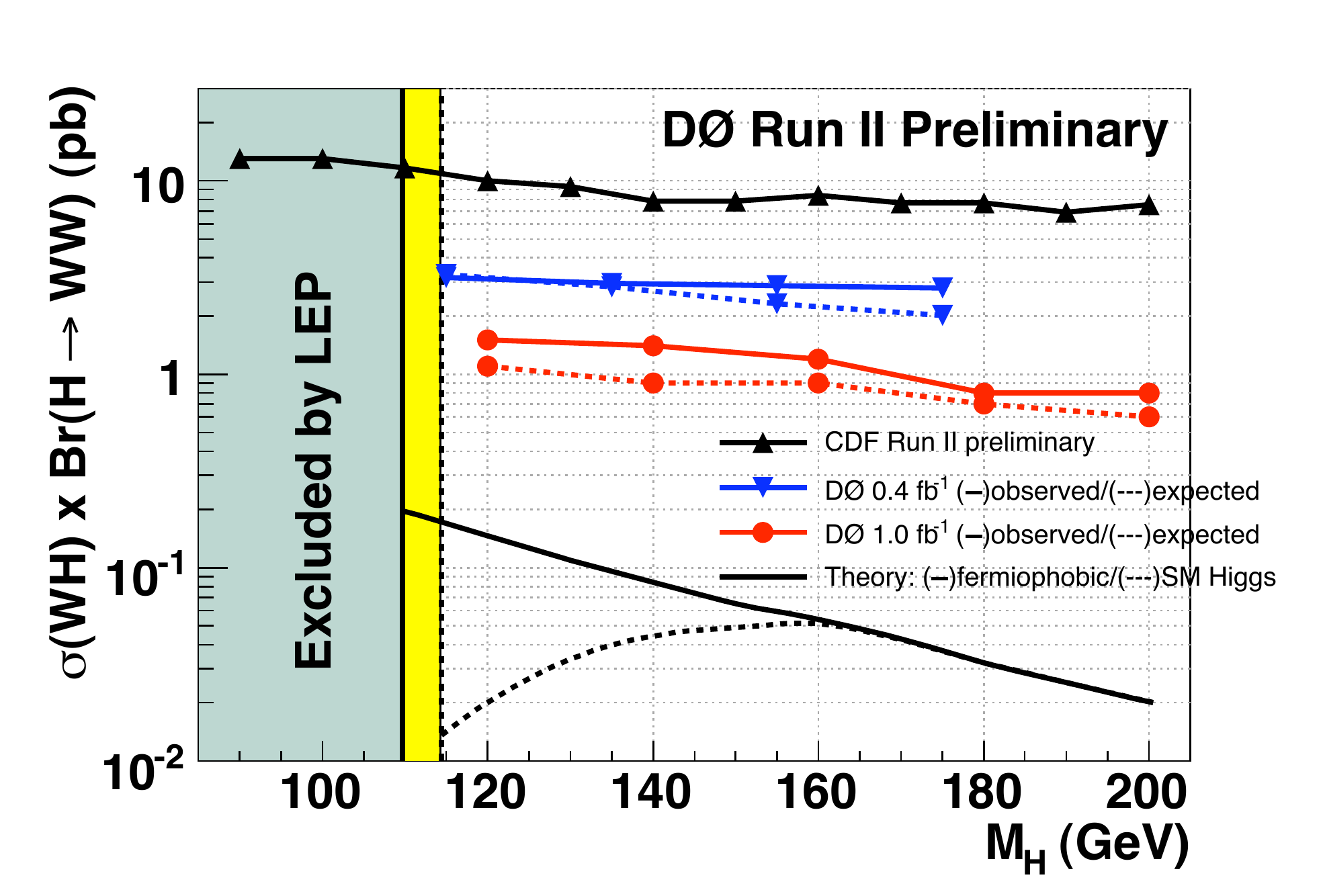}
\caption{\DO search for \WWW.  The three plots on the left show the invariant mass of the dilepton system for data and background prediction (from left to right $ee$, $e\mu$, $\mu\mu$).  The plot on the right shows the limit that has been extracted from those, as well as limits from previous measurement.  The limit is compared to both the SM Higgs prediction and the fermiophobic Higgs prediction.}
\label{p03}
\end{figure*}

\begin{figure*}[tbph]
\centering
\begin{tabular}{|c|c|}
\hline & \textbf{events} \\
\hline bkg & $141.7\pm9.2$ \\
\hline sig($m_H=110$) & $0.938\pm0.050$ \\
\hline sig($m_H=160$) & $0.272\pm0.016$ \\
\hline data & 134\\
\hline
\end{tabular}
\includegraphics[width=60mm]{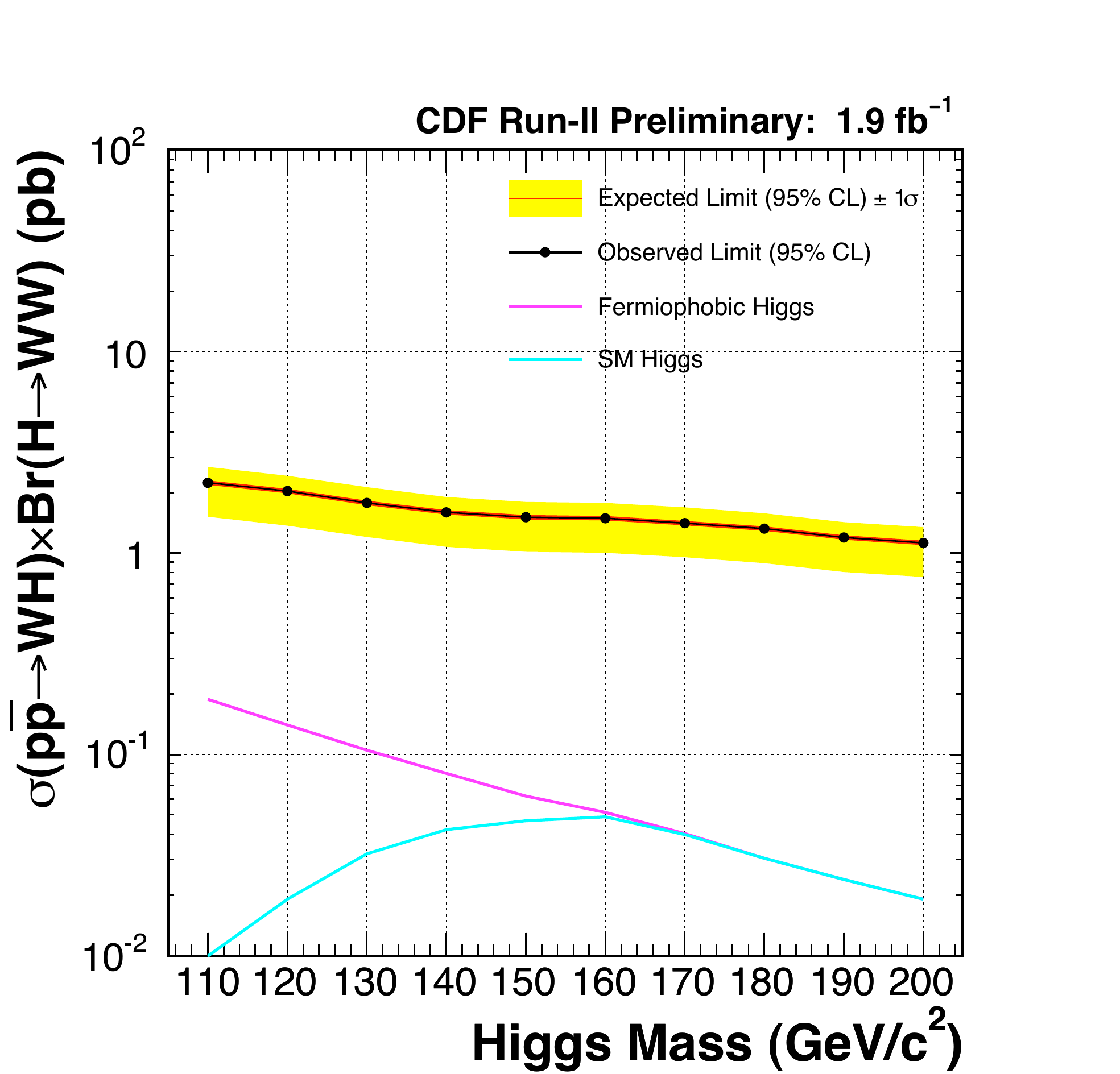}
\includegraphics[width=60mm]{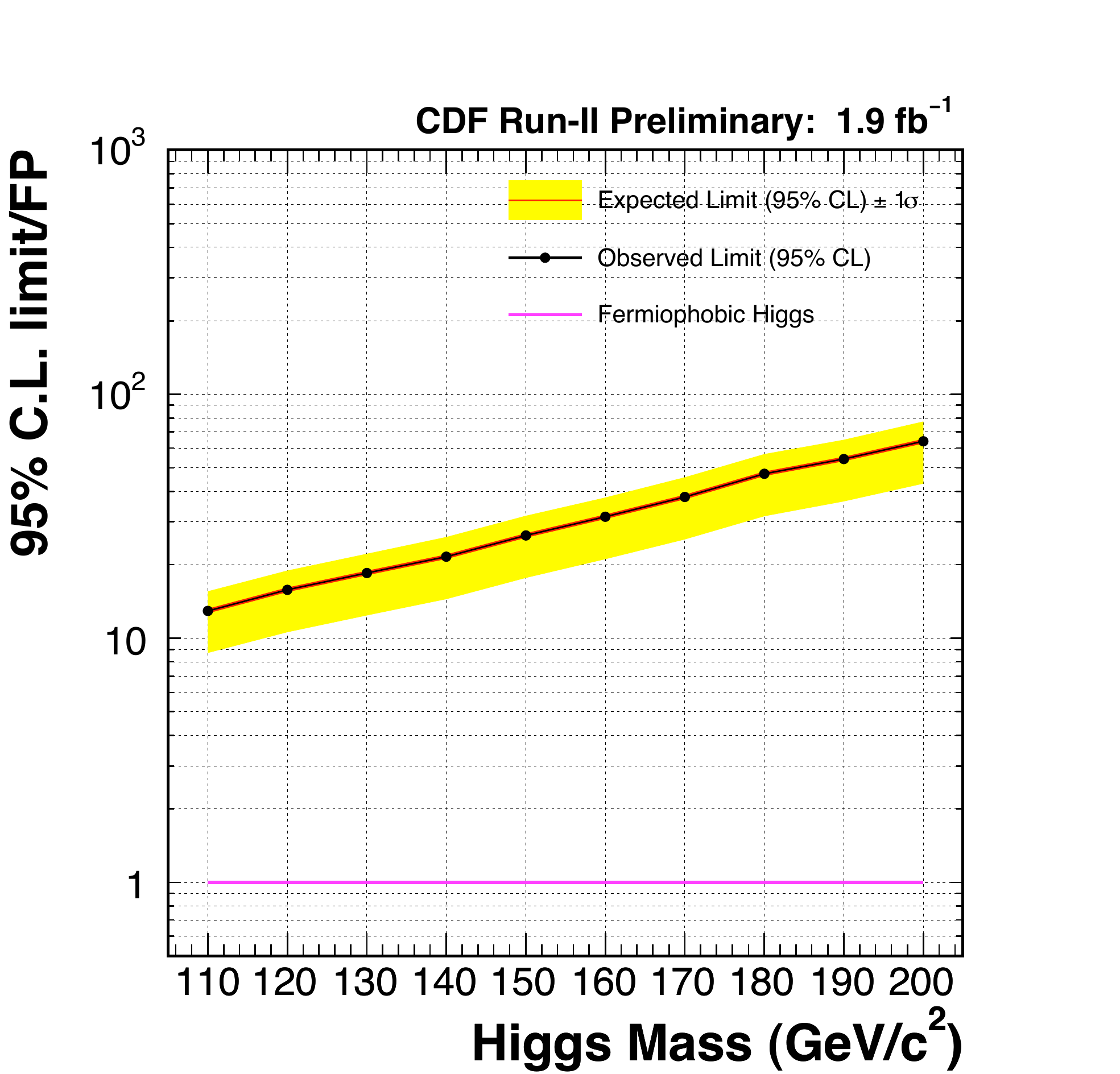}
\caption{CDF search for \WWW.  The table shows the expected and observed events.  The left plot shows the limit on the cross section, extracted from the event counts.  The limit is compared to the theory prediction for SM Higgs and fermiophobic Higgs.  The plot on the right shows the limit normalized to the fermiophobic Higgs prediction.}
\label{p04}
\end{figure*}

The \DO search for \WWW~\cite{DO} looks for two same sign leptons and vetos on a third lepton.  The sample is then divided according to lepton types.  The final limit is extracted from the dilepton invariant mass distribution.  The results can be seen in figure~\ref{p03}.  The CDF search for \WWW~\cite{CDF} also looks for exactly two leptons of the same sign and divides the sample by lepton type, but it performs a simple counting experiment instead of extracting the limit from the dilepton invariant mass.  The results can be seen in figure~\ref{p04}.

\begin{figure*}[tbph]
\centering
\includegraphics[width=\widthb]{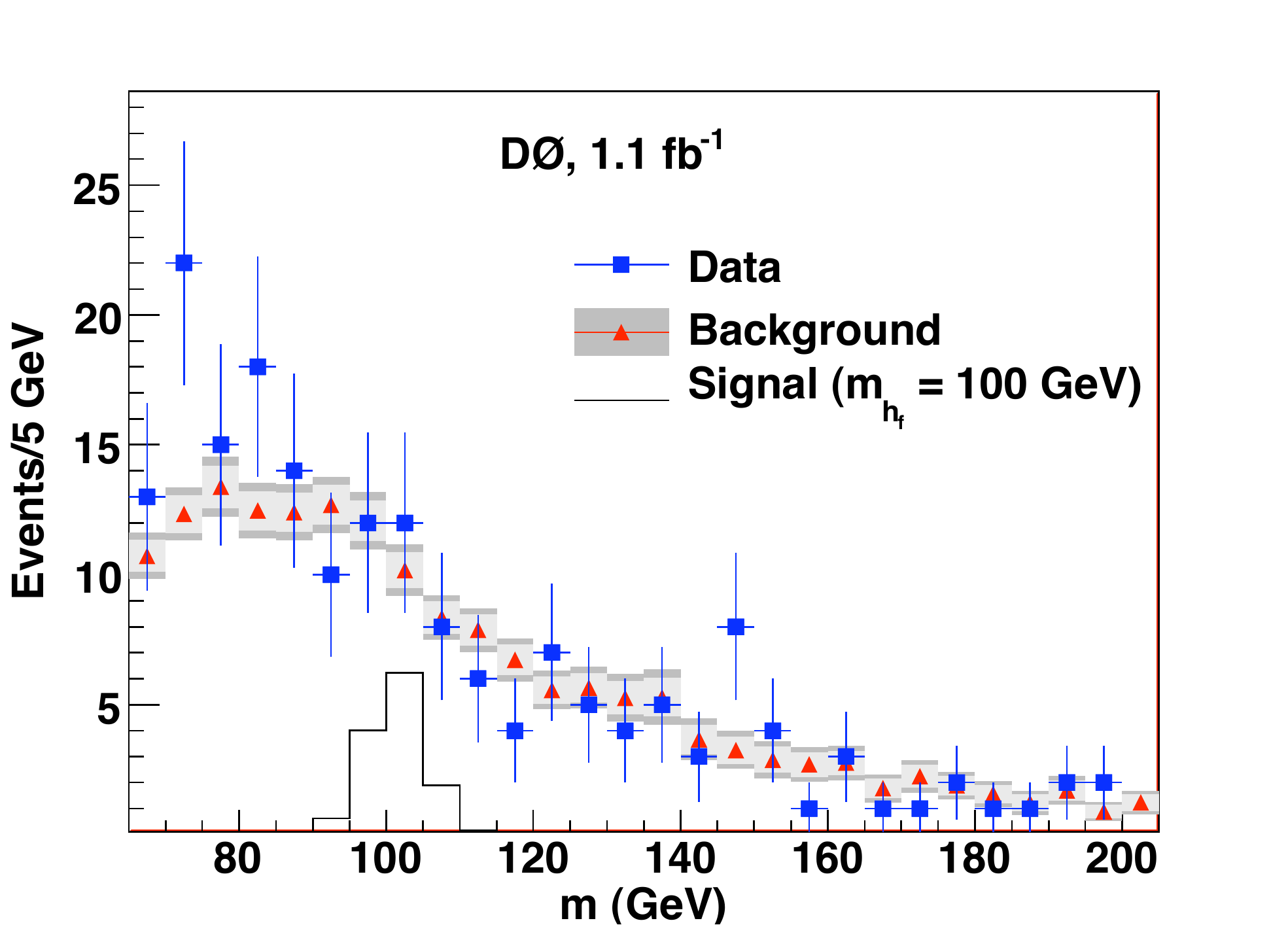}
\includegraphics[width=\widthb]{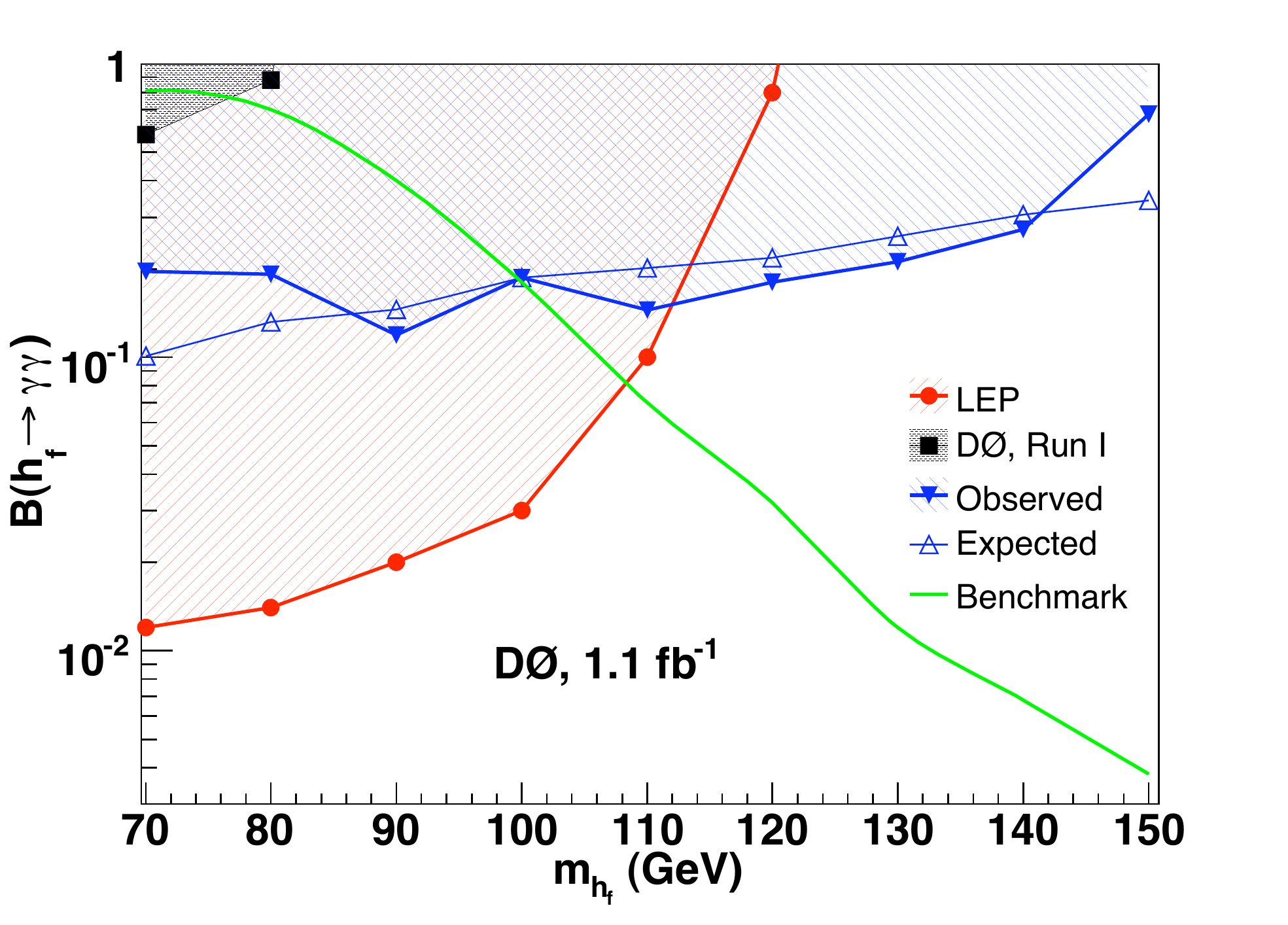}
\caption{\DO search for \dipho.  The plot on the left shows the diphoton mass spectrum.  It compares the observed data to the background prediction and the signal peak for a Higgs mass of 100\ GeV.  The plot on the right shows the expected and observed limit on the branching ratio of higgs to diphotons, and compares it to previous measurements by D0 and LEP.}
\label{p05}
\end{figure*}

\DO performed a search for \dipho~\cite{DO}.  That search employs a neural network to distinguish between photons and jets.  The diphoton mass spectrum is then scanned for an excess and a mass dependent limit is set.  The results can be seen in figure~\ref{p05}.

\begin{figure*}[tbph]
\centering
\begin{tabular}{|c|c|}
\hline & \textbf{events} \\
\hline SM bkg & $1.1\pm0.2$ \\
\hline data & 0 \\
\hline
\end{tabular}
\includegraphics[width=45mm]{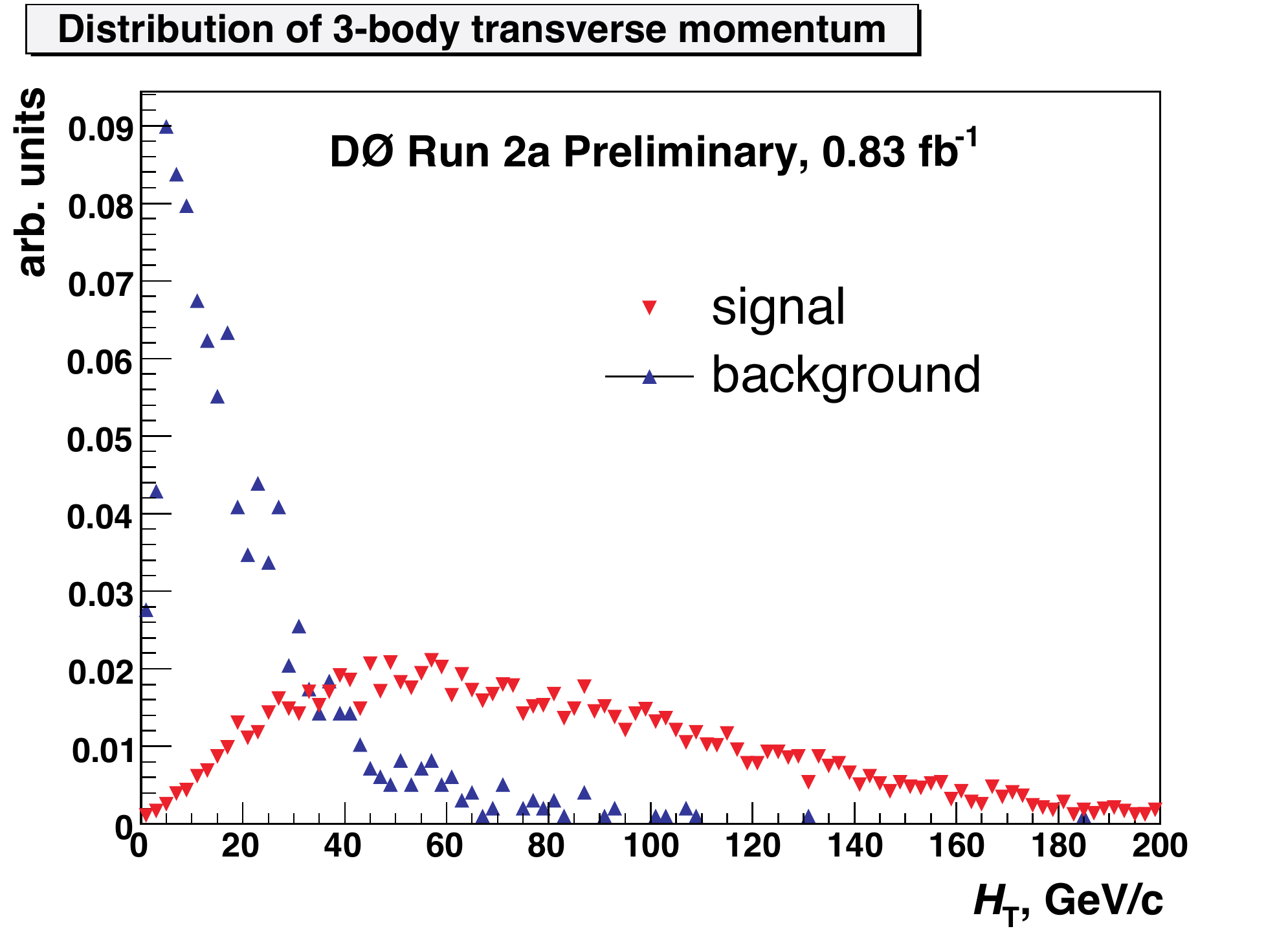}
\includegraphics[width=90mm]{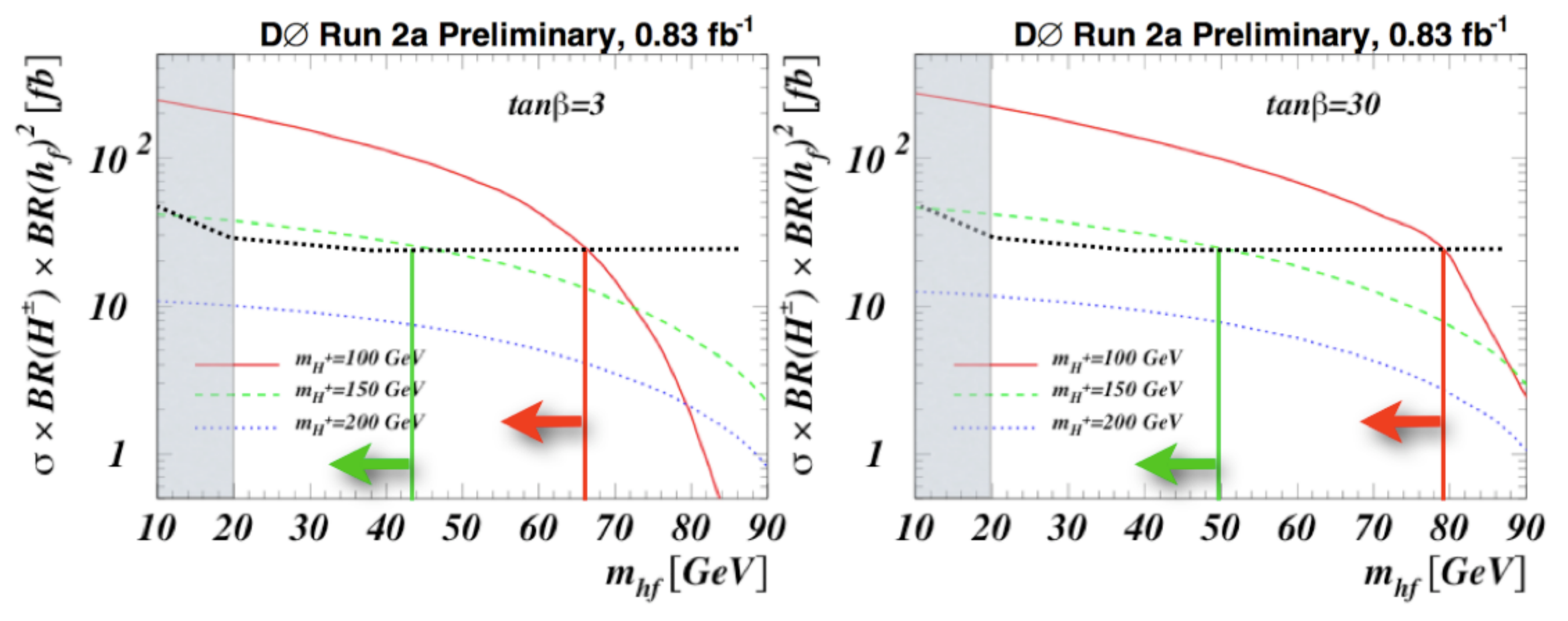}
\caption{\DO search for \tripho.  The signal and background predictions for the total transverse momentum are shown in the plot on the left, the predicted and observed events after applying a cut $H_T>30\,\hbox{GeV}$ are shown in the table.  From these model and mass dependent limits are extracted and shown in the two plots on the right.} \label{p06}
\end{figure*}

\DO also performed a search for \tripho~\cite{DO}.  The basic idea is that the vertex $h_fVV$ is suppressed and instead the vertex $h_fVH$ is employed, where $H$ is another Higgs boson.  The search only looks for three photons, allowing one photon to be lost.  A cut on the transverse momentum of the photons is employed to separate out the background, and the remaining events are treated as a counting experiment.  The results can be seen in figure~\ref{p06}.

\section{SUSY Higgs Searches}

The SUSY Higgs searches exploit the fact that at high $\tan\beta$ the coupling of the Higgs to bottom quark and tau leptons is enhanced with $\tan^2\beta$, which has a number of effects that are exploited in the search.  The Higgs cross section increases from femtobarns to picobarns.  The Higgs is often produced by the fusion of two bottom quarks producing a final state with two extra bottom quarks, which can be tagged to reduce backgrounds.  Furthermore the Higgs decay changes and is in 90\% of cases a pair of bottom quarks and in the remaining 10\% into a pair of tau leptons.

\begin{figure*}[tbph]
\centering
\includegraphics[width=\widthc]{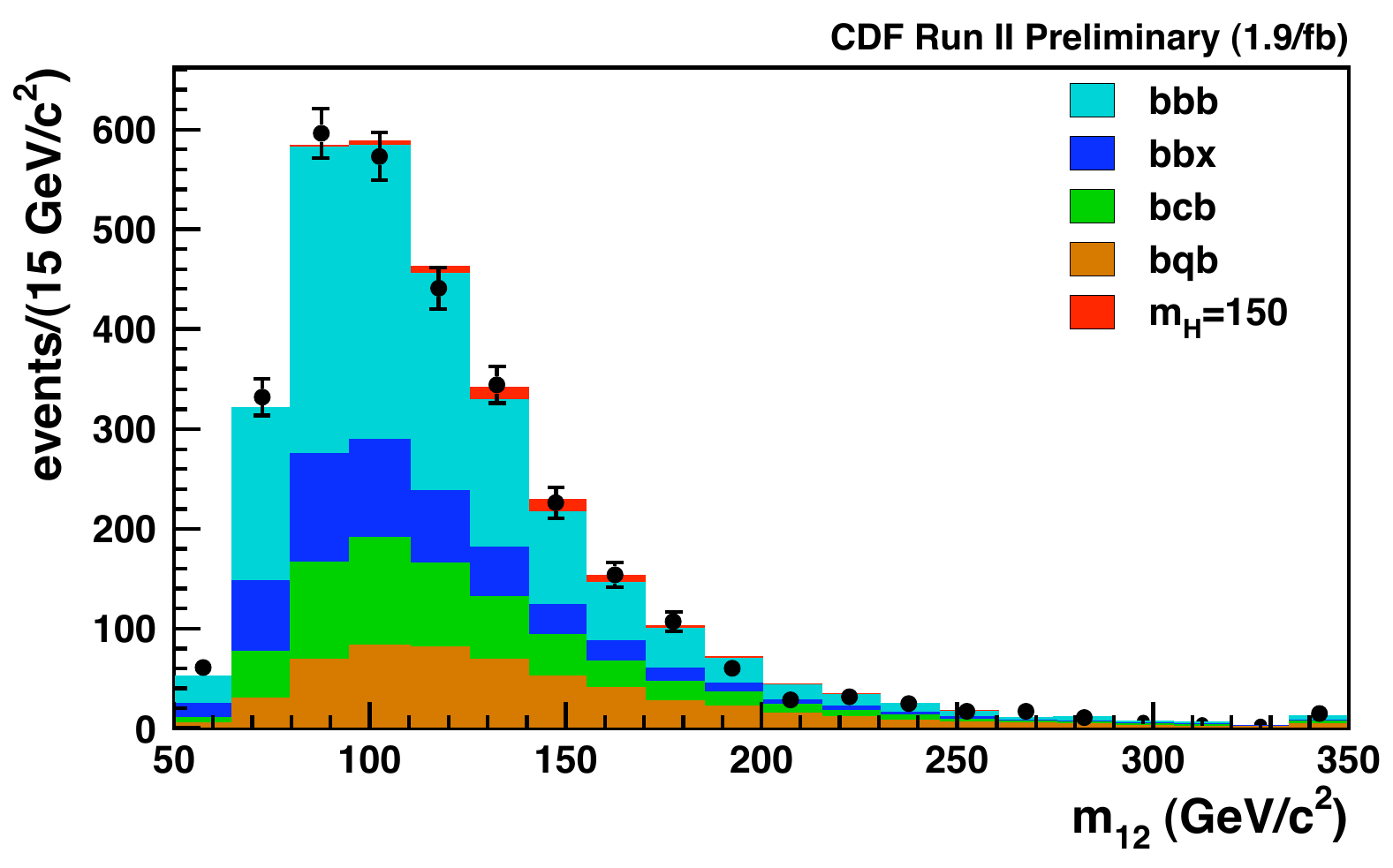}
\includegraphics[width=\widthc]{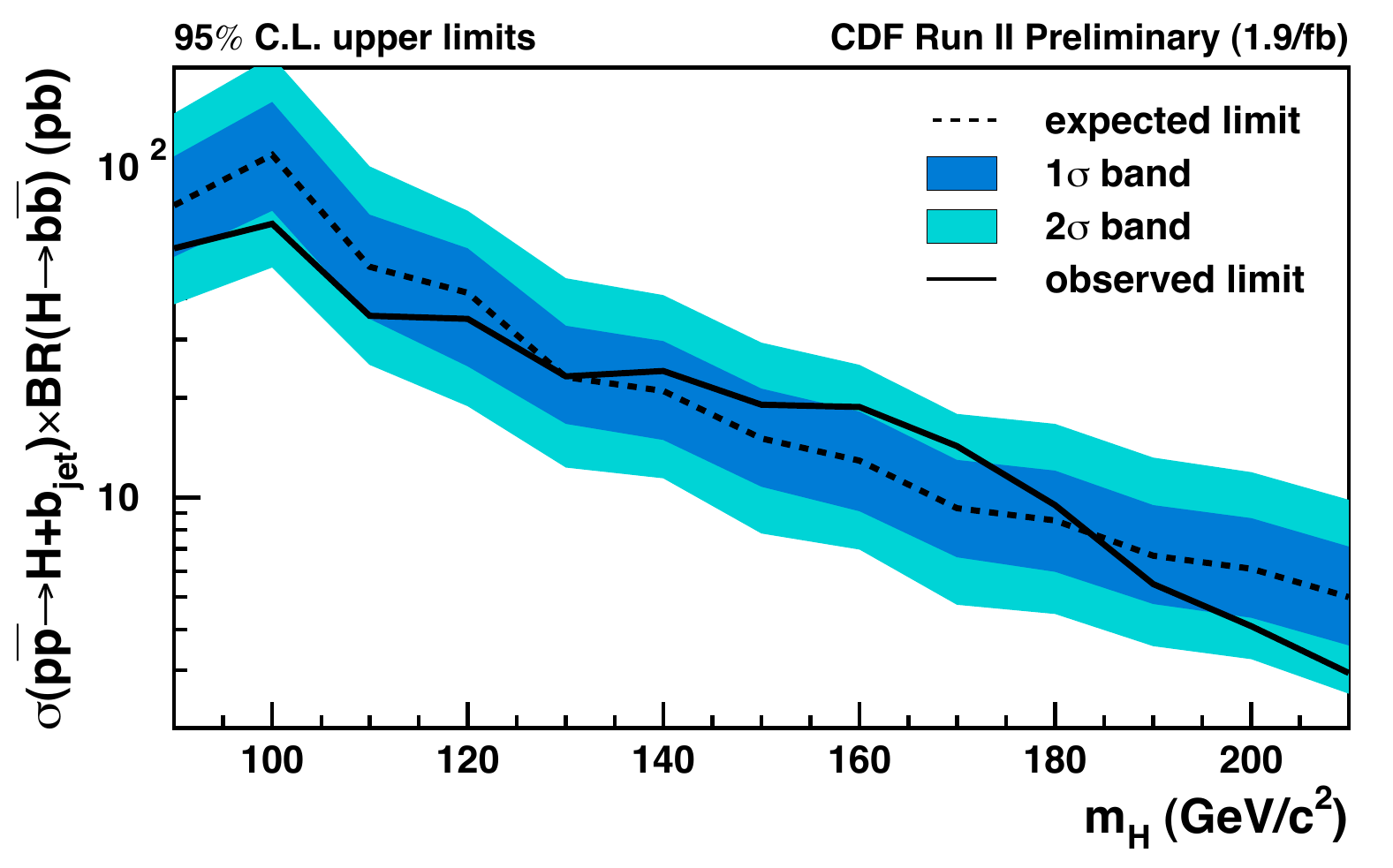}
\includegraphics[width=\widthc]{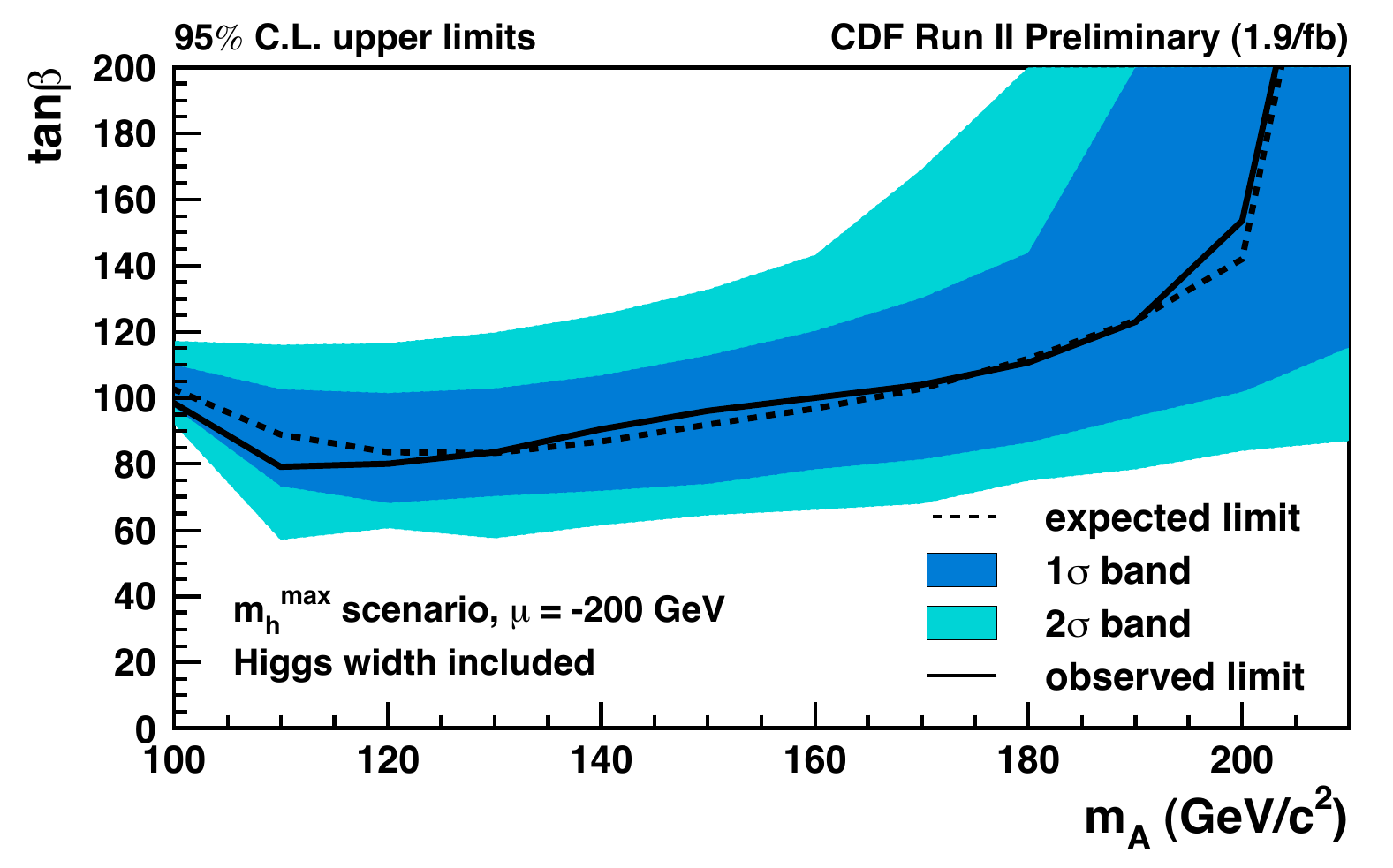}
\caption{CDF search for \bbb.  The invariant mass spectrum for signal and background predictions are compared in the plot on the left.  The extracted mass dependent cross section limit is shown in the middle plot.  The $\tan\beta$ limit is shown on the right (taking Higgs width effects into account).} \label{p09}
\end{figure*}

CDF performed a search for \bbb~\cite{CDF}, where the choice of one associated bottom jet was due to the better signal to background ratio compared to zero or two associated bottom jets.  In the search the two and three jet distributions are fit to obtain the signal and background distributions.  From this, mass dependent cross section limits are obtained, and $\tan\beta$ limits which take the Higgs width effects at high $\tan\beta$.  The results can be seen in figure~\ref{p09}.

\begin{figure*}[tbph]
\centering
\includegraphics[width=\widthc]{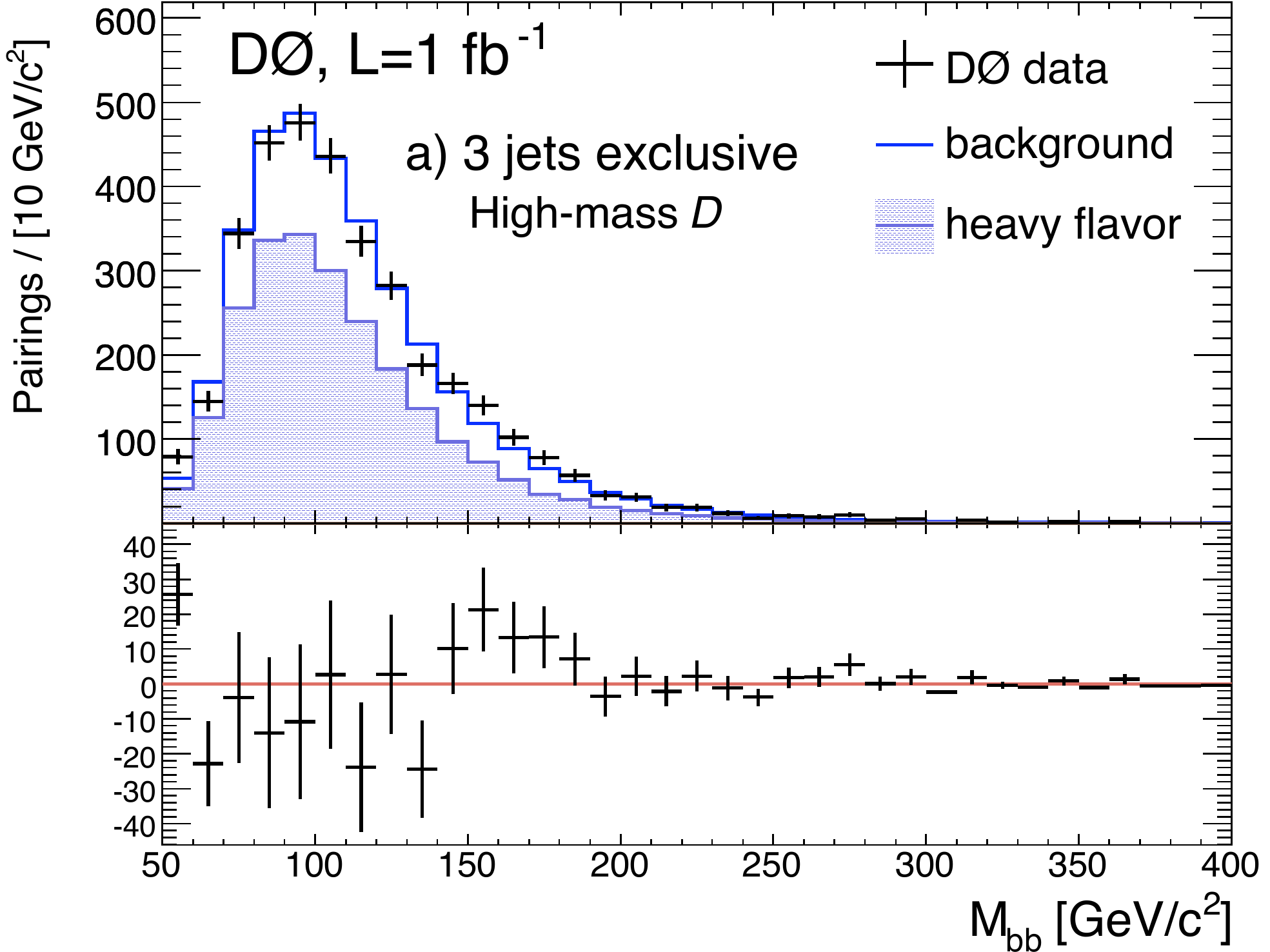}
\includegraphics[width=\widthc]{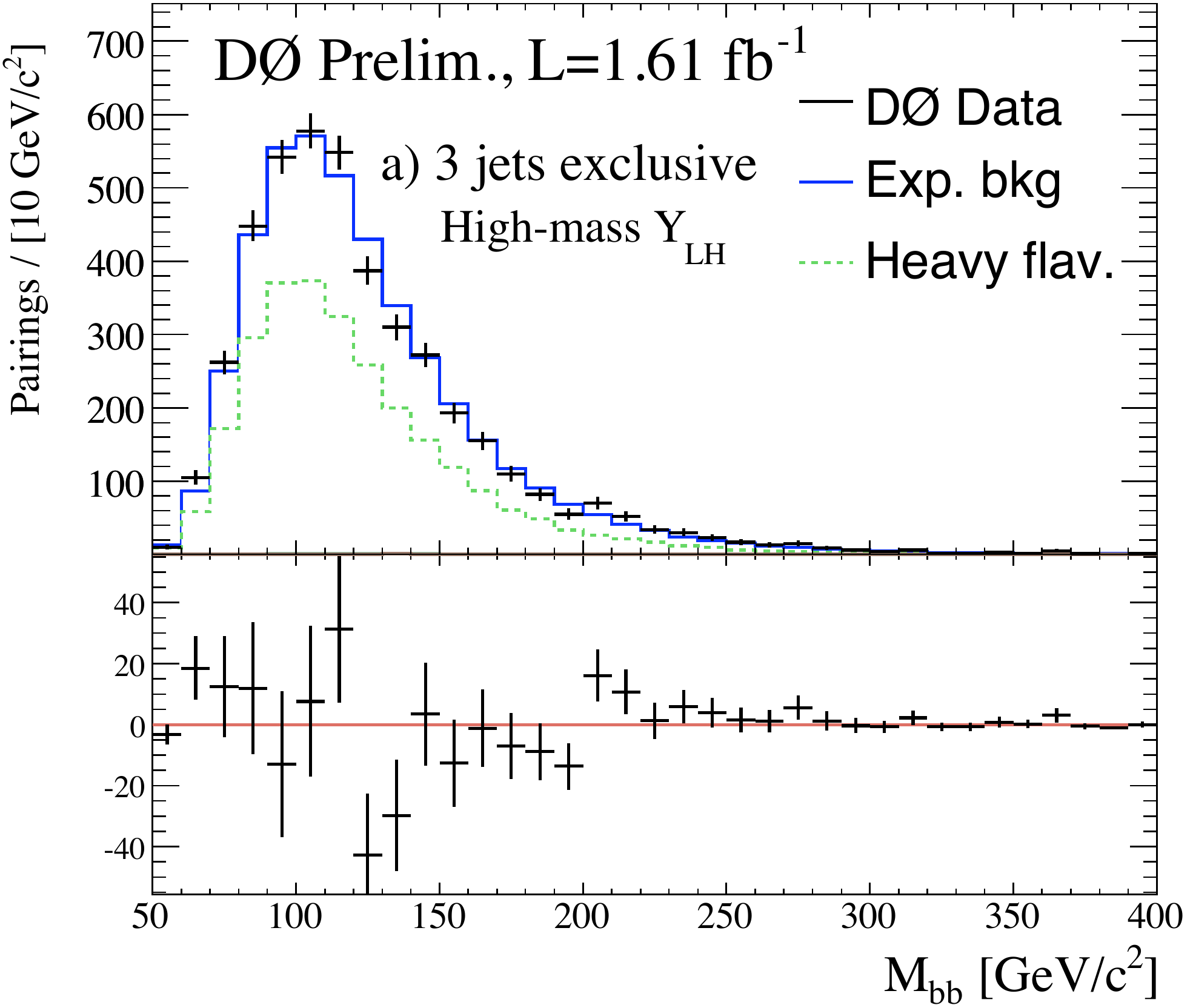}
\includegraphics[width=\widthc]{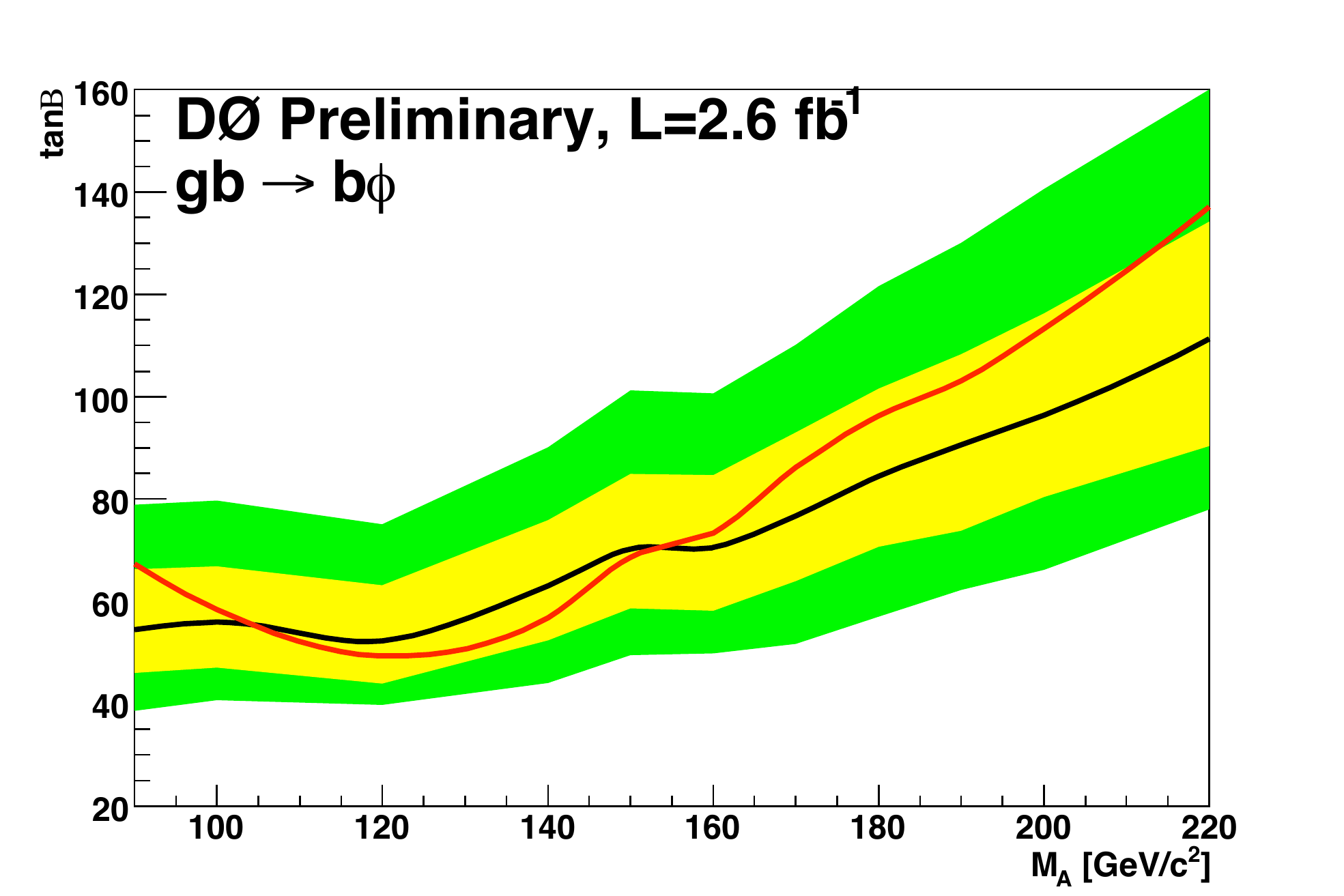}
\caption{\DO search for \bbb.  The invariant mass spectrum for two subchannels is shown in the two left plots.  The extracted mass dependent $\tan\beta$ limits are shown on the right.} \label{p10}
\end{figure*}

\DO also performed a search for \bbb~\cite{DO}.  For the search the data is split into three, four and five jet bins.  Within these bins the two and three jet distributions are fit to obtain the signal and background contributions.  From these the mass dependent $\tan\beta$ limit is extracted.  The results can be seen in figure~\ref{p10}.

\begin{figure*}[tbph]
\centering
\includegraphics[width=\widthdb]{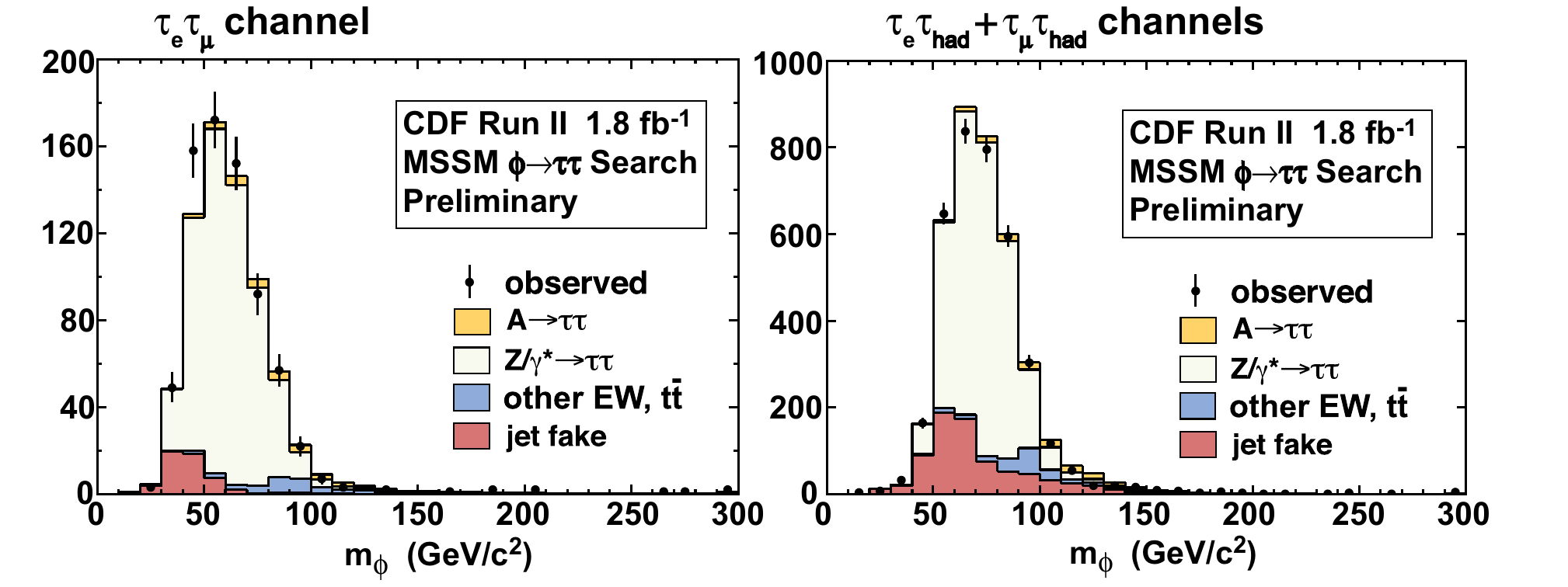}
\includegraphics[width=\widthd]{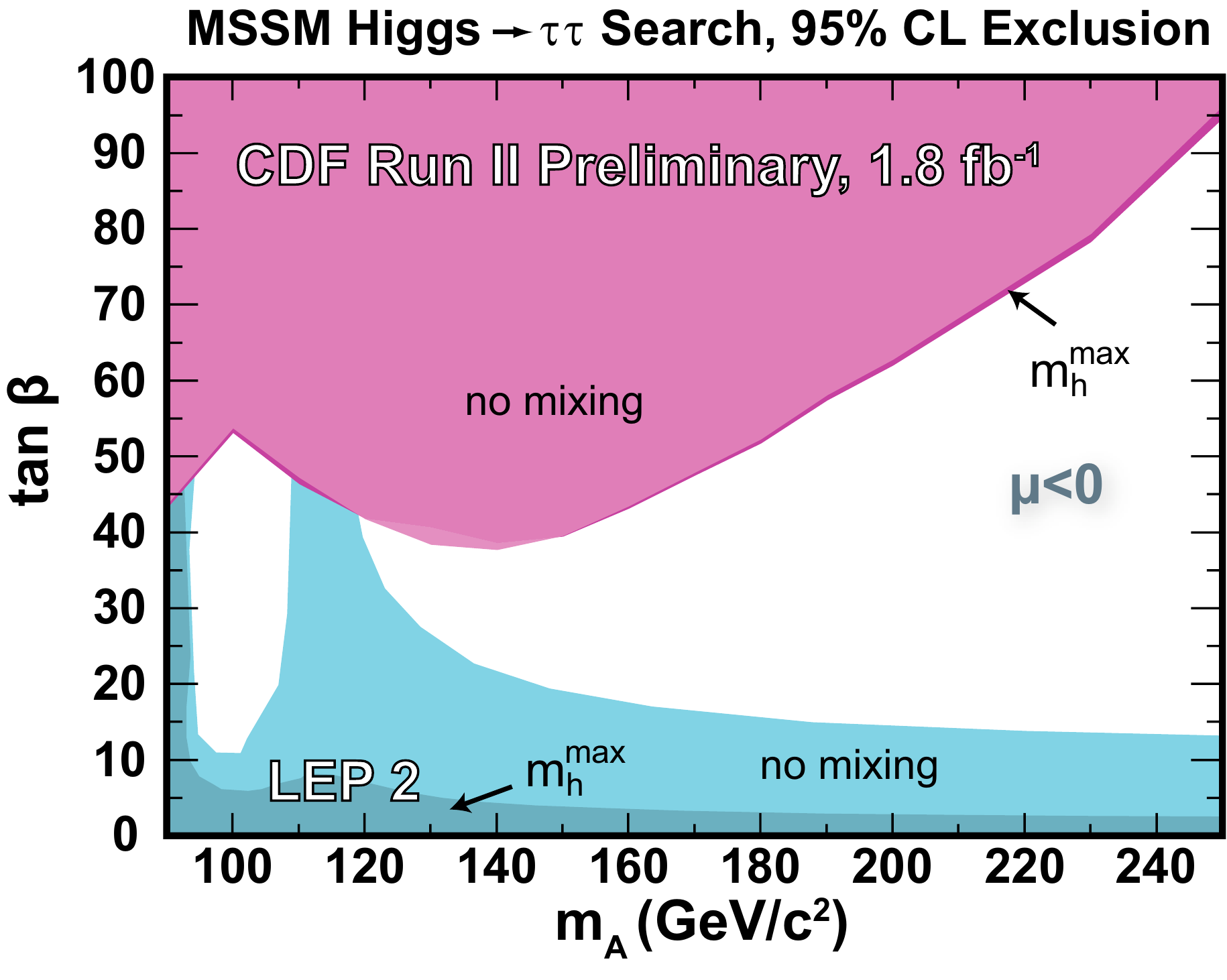}
\includegraphics[width=\widthd]{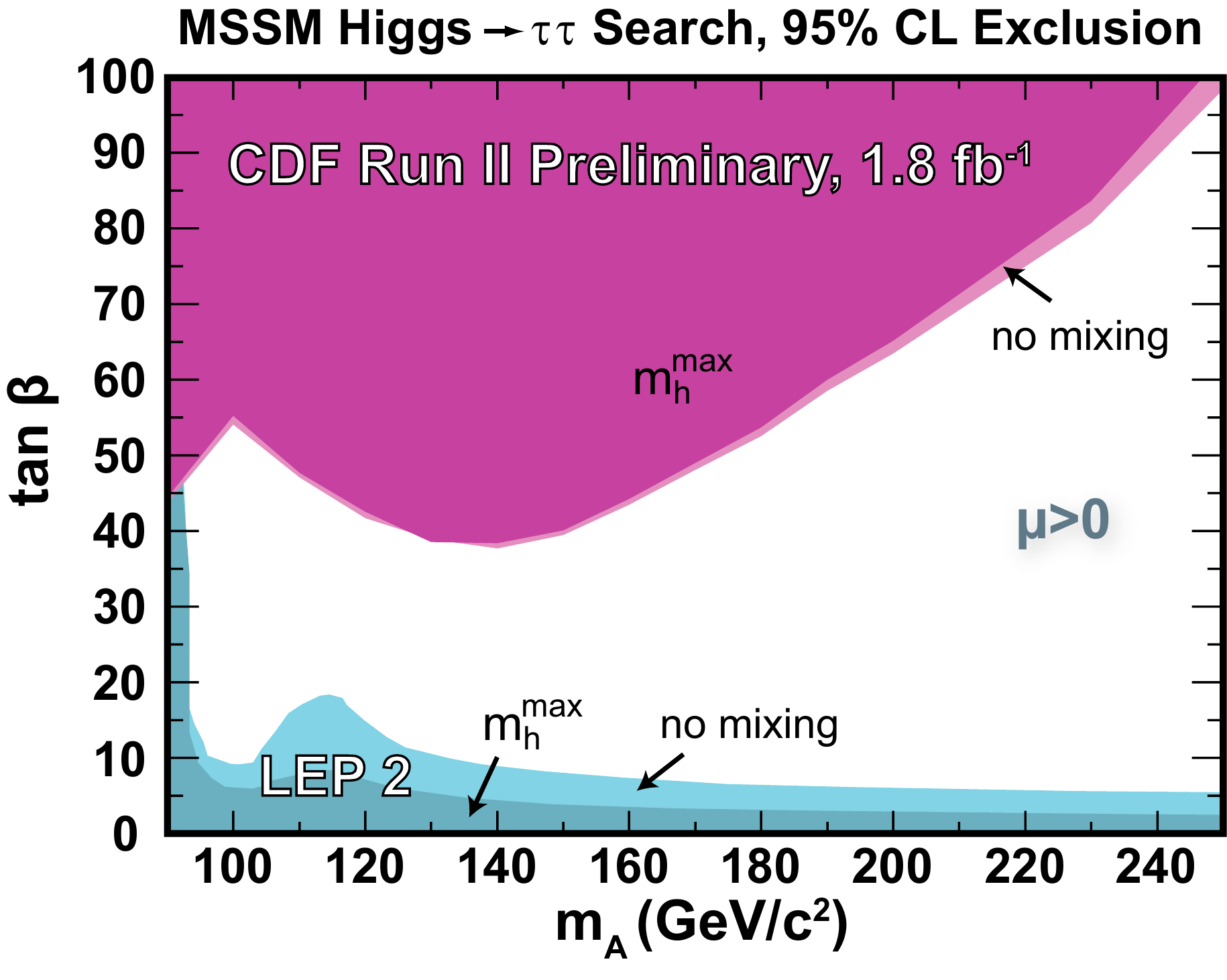}
\caption{CDF search for \ditau.  The plots on the left show the invariant mass spectrum for two of the sub channels.  The plots on the right show the model and mass dependent limits.} \label{p11}
\end{figure*}

CDF performed a search for \ditau~\cite{CDF}.  Due to the higher purity of the ditau sample no additional bottom jets are required.  The sample is split according to the tau decay mode.  The signal template is fit to the ditau visible mass spectrum to obtain the mass and model dependent $\tan\beta$ limits.  The results can be seen in figure~\ref{p11}.

\begin{figure*}[tbph]
\centering
\includegraphics[width=\widthc]{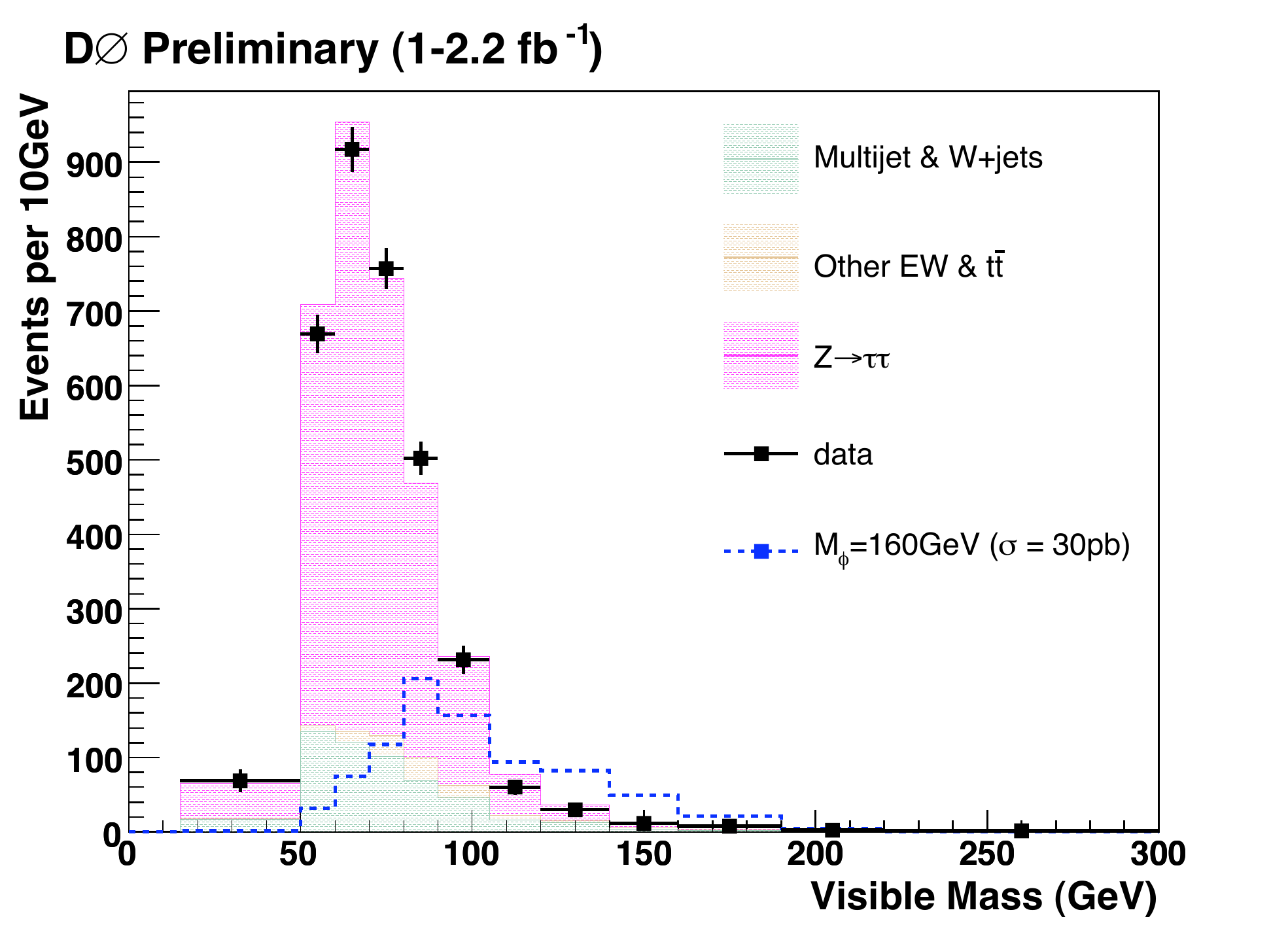}
\includegraphics[width=\widthc]{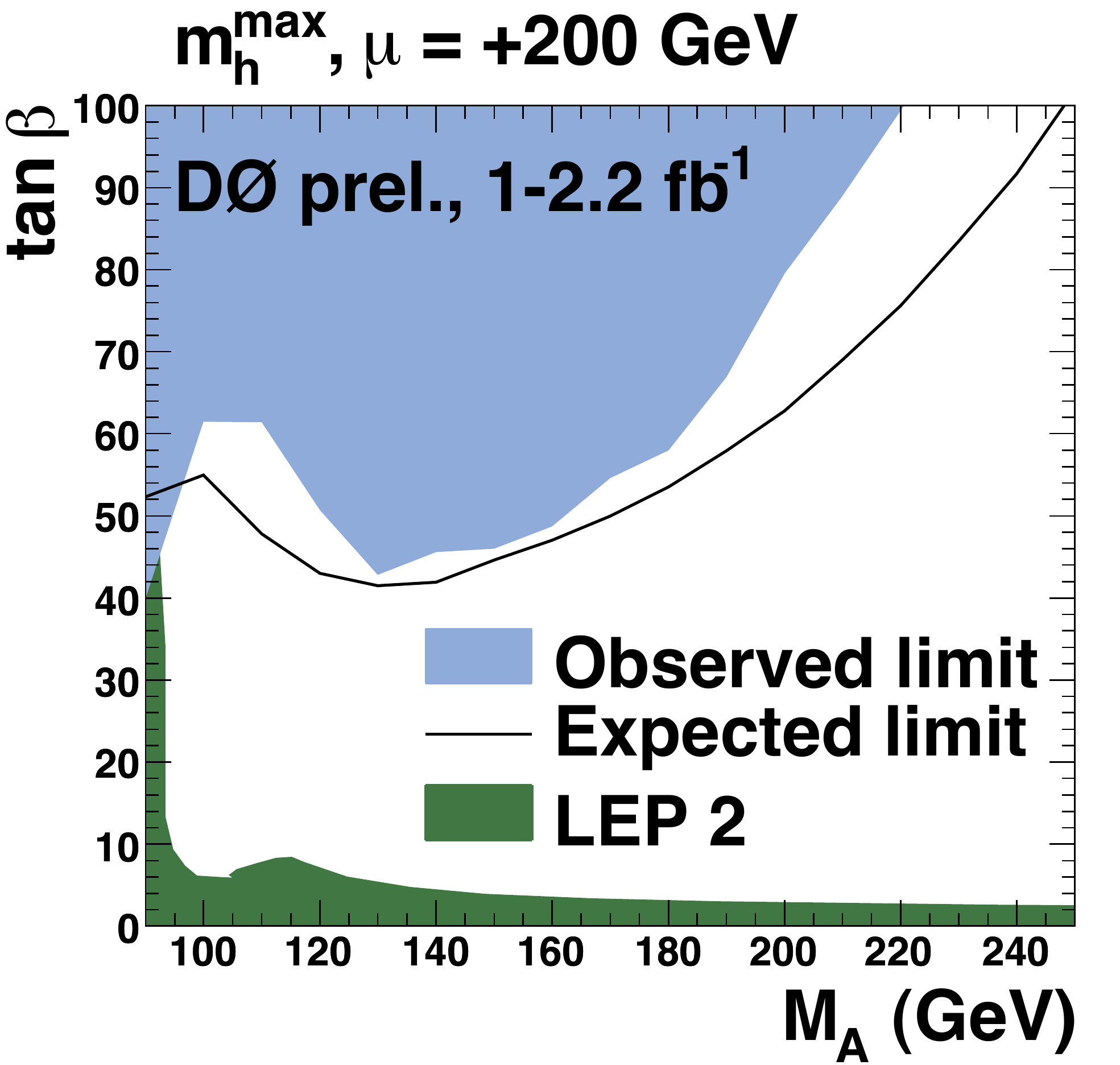}
\includegraphics[width=\widthc]{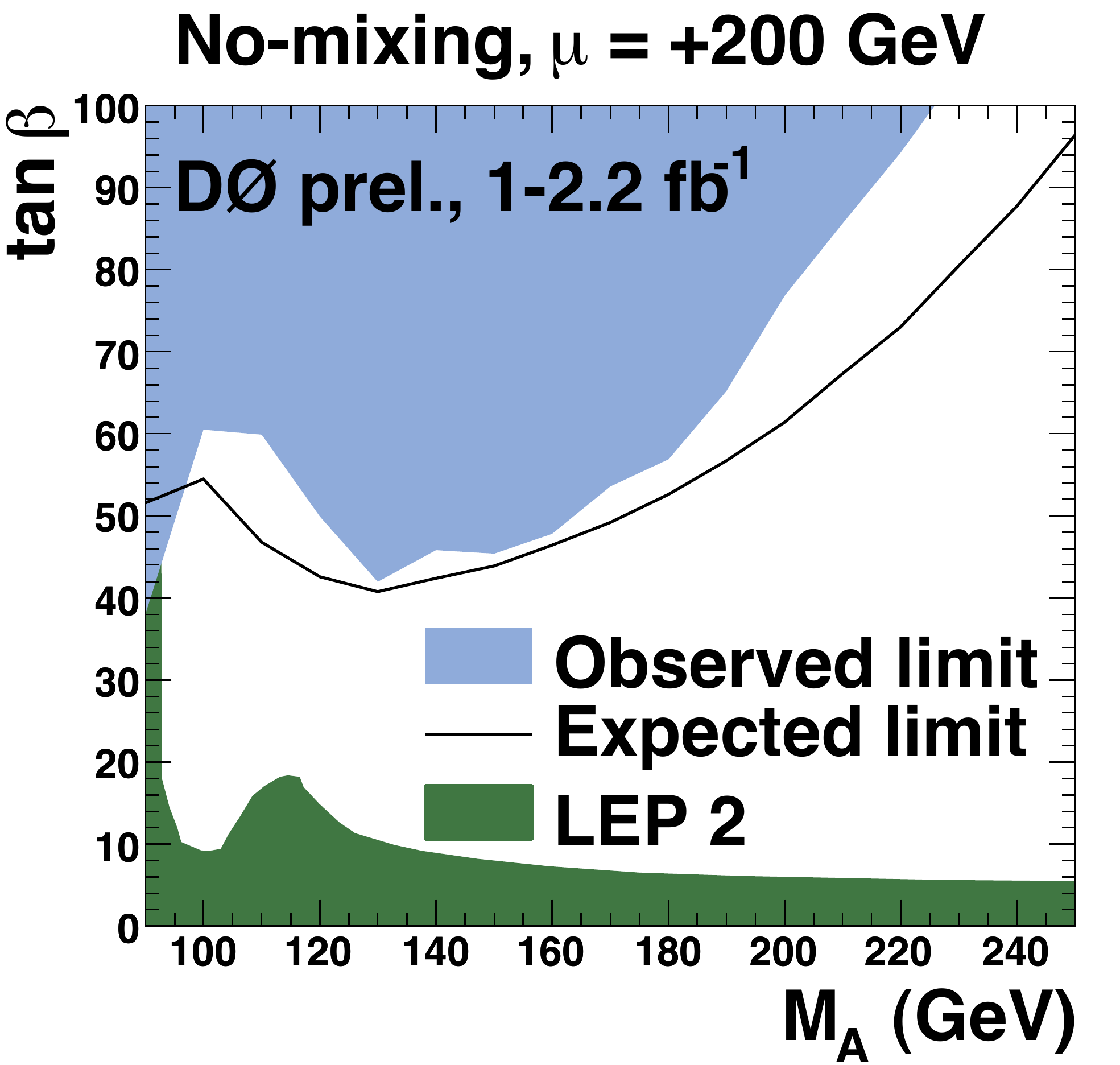}
\caption{\DO search for \ditau.  The plot on the left compares signal and background prediction to data for one subchannel.  The two plots on the right show the extracted model and mass dependent limits on $\tan\beta$.} \label{p12}
\end{figure*}

\DO performed two searches for Higgs decaying into a tau pair~\cite{DO}.  The one search is in the \ditau channel.  For this the sample is divided by tau decay type and run period.  The signal template is then fit to the visible mass spectrum to extract the model and mass dependent $\tan\beta$ limit.  The results can be seen in figure~\ref{p12}.

\begin{figure*}[tbph]
\centering
\begin{tabular}{|c|c|c|c|}
\hline $\tau_h$ type & I & II & III \\
\hline bkg & $1.2\pm0.2$ & $2.6\pm0.3$ & $2.5\pm0.2$\\
\hline signal & $0.6\pm0.1$ & $3.5\pm0.5$ & $1.2\pm0.2$\\
\hline data & 0 & 1 & 2 \\
\hline
\end{tabular}
\includegraphics[width=60mm]{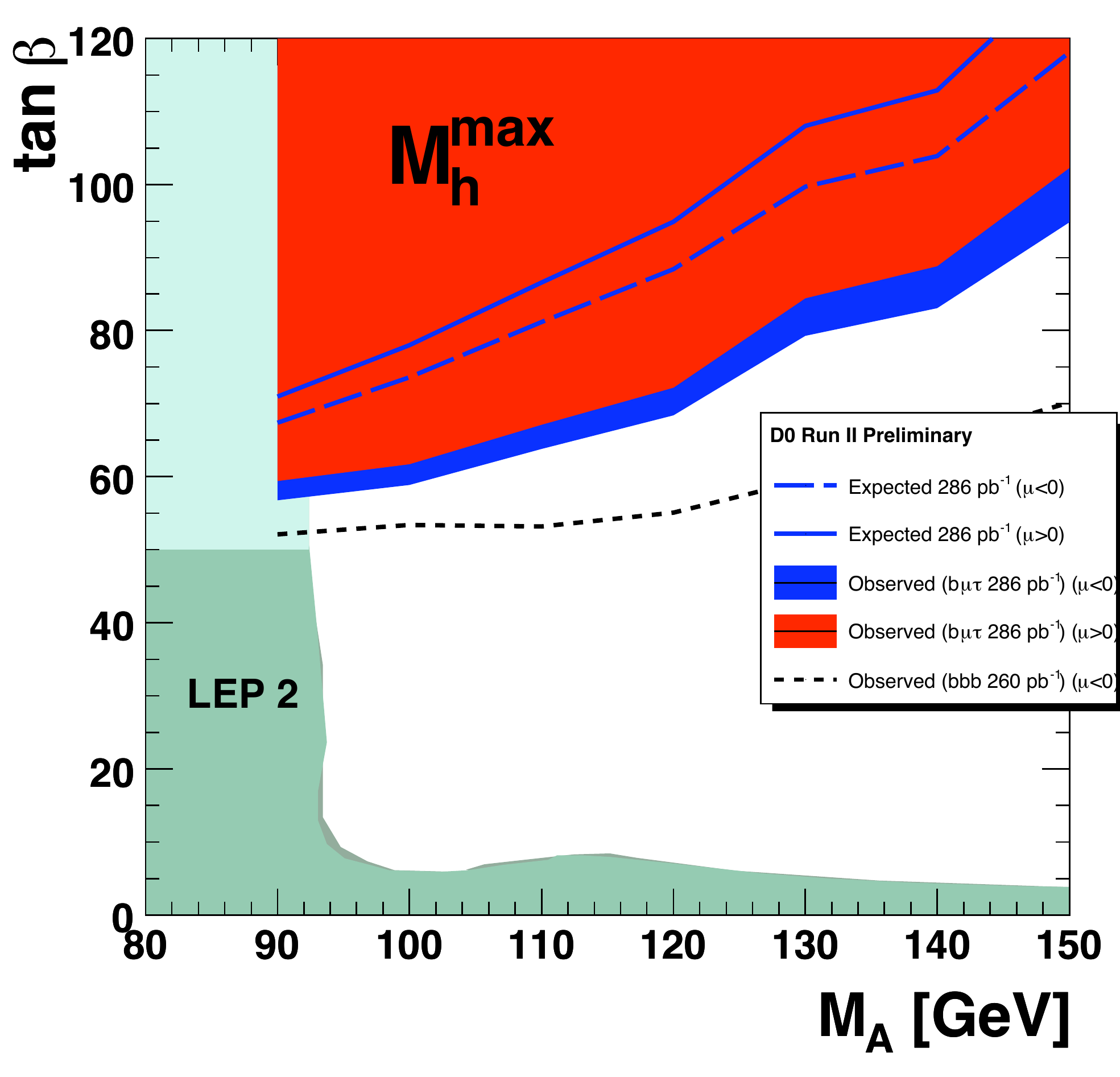}
\includegraphics[width=60mm]{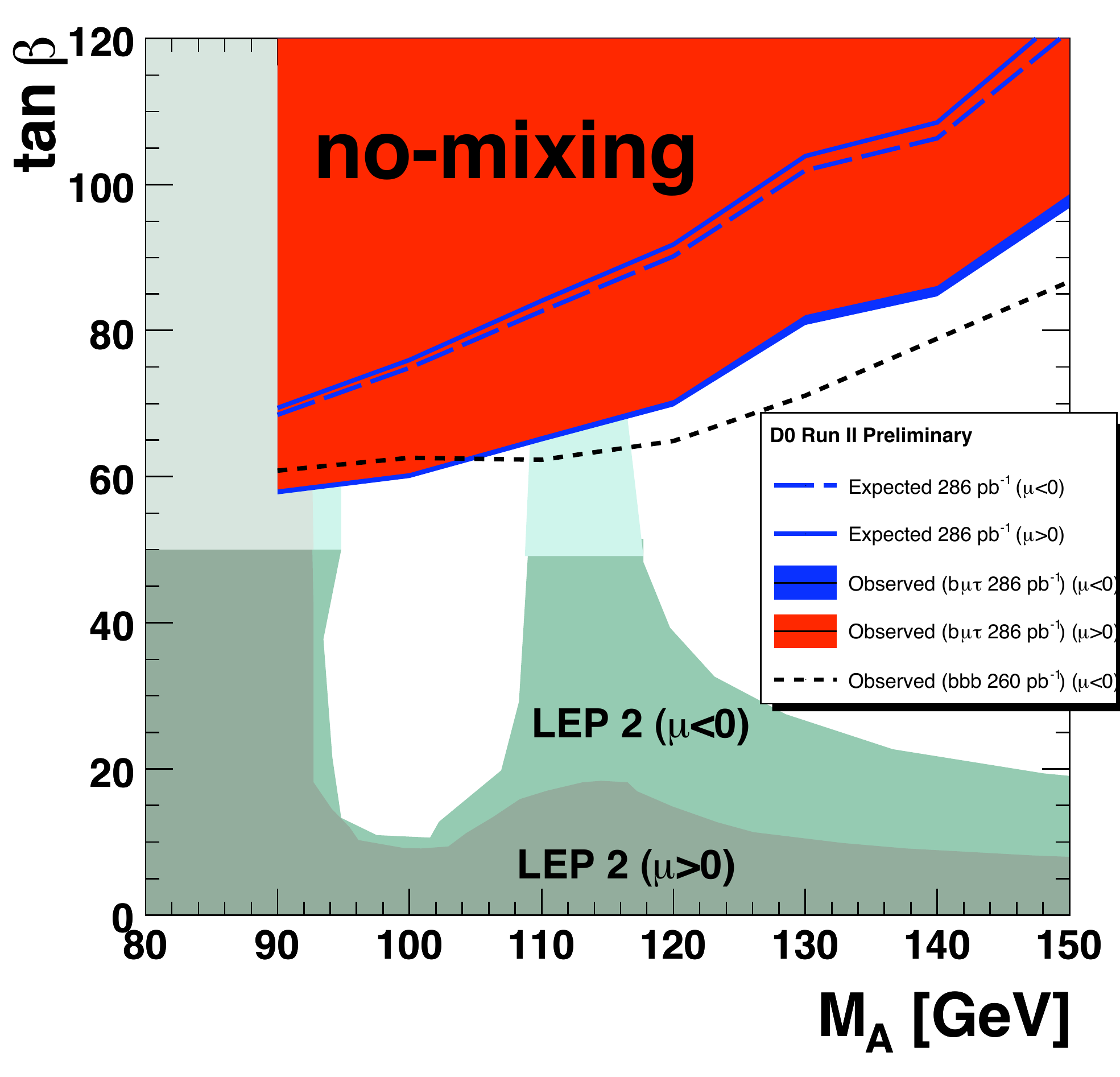}
\caption{\DO search for \bditau.  The table shows the predicted and observed events in the different subchannels.  The plots show the extracted mass and model dependent limits on $\tan\beta$} \label{p13}
\end{figure*}

The other search is in the \bditau search, which tags an additional bottom quark to reduce the background rate.  One tau is required to decay to a muon, while the other is required to decay hadronically.  The sample is then divided according to the hadronic tau decay type.  In each decay category a counting experiment is performed to extract the model and mass dependent $\tan\beta$ limits.  The results can be seen in figure~\ref{p13}.

\section{Summary and Outlook}

The CDF and \DO collaborations have performed a large number of searches for signs of a Fermiophobic or SUSY Higgs.  In the absence of signal, limits have been set on the various models.  For the near future, updates to several of these searches can be expected, as well as a combination of these searches into an overall Tevatron limit.

\end{document}